\documentclass[%
 reprint,
nofootinbib,
 amsmath,amssymb,
 aps,
prl,
]{revtex4-1}

\usepackage{graphicx}% Include figure files
\usepackage{bm}% bold math
\usepackage{epsfig}
\usepackage{color}
\usepackage{slashed}
\usepackage[
	colorlinks=true,
	urlcolor=blue,
	linkcolor=red,
	citecolor=blue
]{hyperref}
\usepackage{cleveref}
\crefname{figure}{fig.}{figs.} 
\Crefname{figure}{Fig.}{Figs.}
\crefname{equation}{eq.}{eqs.} 
\Crefname{equation}{Eq.}{Eqs.}

\begin{document}

\title{Let there be Light Dark Matter:\\ The gauged $U(1)_{L_\mu-L_\tau}$ case}

\author{Patrick Foldenauer}
\email[Email: ]{foldenauer@thphys.uni-heidelberg.de}
\affiliation{%
Institut f\"ur Theoretische Physik, Universit\"at Heidelberg, Philosophenweg 16, 69120 Heidelberg, Germany
}%

\date{February 12, 2019}% It is always \today, today,

\begin{abstract}
As experimental null results increase the pressure on heavy weakly interacting massive particles (WIMPs) as an explanation of thermal dark matter (DM), it seems timely to explore previously overlooked regions of the WIMP parameter space. In this work we extend the minimal gauged $U(1)_{L_\mu-L_\tau}$ model studied in \cite{Bauer:2018onh} by a light (MeV-scale) vector-like fermion $\chi$. Taking into account constraints from cosmology, direct and indirect detection we find that the standard benchmark of $M_V=3 m_\chi$ for DM coupled to a vector mediator is firmly ruled out for unit DM charges. However, exploring the near-resonance region $M_V\gtrsim 2 m_\chi$ we find that this model can simultaneously explain the DM relic abundance $\Omega h^2 =0.12$ and the $(g-2)_\mu$ anomaly. Allowing for small charge hierarchies of $\lesssim\mathcal{O}(10)$, we identify a second window of parameter space in the few-GeV region, where $\chi$ can account for the full DM relic density.
\end{abstract}

\maketitle

%%%%%%%%%%%%%%%%%%%%%%%%%%%%%%%
\section{Introduction}\label{sec:intro}
%%%%%%%%%%%%%%%%%%%%%%%%%%%%%%%

The advent of the Standard Model of particle physics (SM) 
 \cite{Glashow:1961tr,Weinberg:1967tq,Salam:1968rm} was one of the biggest milestones in physics. The excellent agreement of its predictions with data have allowed physicists to embark on an era of precision studies of physics at the smallest scales accessible. Nevertheless, there are some hints for physics beyond the Standard Model (BSM), one of the most intriguing ones being the $\sim 4 \sigma$ excess of the anomalous muon magnetic moment $(g-2)_\mu$ measured at the BNL E821 experiment    \cite{Bennett:2006fi,Jegerlehner:2009ry,Davier:2010nc,Hagiwara:2011af}. The upcoming E989 experiment at Fermilab aims at a fourfold improvement in the experimental sensitivity compared to  E821 \cite{Grange:2015fou} thereby potentially pushing the significance above $5\sigma$, if the excess is due to BSM physics. \par
 Possibly one of the biggest shortcomings of the SM, however, is the absence of a viable candidate for DM \cite{Zwicky:1933gu}.
Even in light of the recent determination of the cosmic DM abundance with unprecedented accuracy to $\Omega_\text{DM}h^2=0.120\pm0.001$  \cite{Aghanim:2018eyx} the true nature of DM remains unknown. Pinning down the exact properties of DM has inspired a myriad of particle physics models. One particularly well-studied class of DM candidates are WIMPs (see \cite{Arcadi:2017kky} for a recent review). Rather stringent bounds on heavy WIMPs \cite{Akerib:2016vxi,Cui:2017nnn,Aprile:2018dbl} have lead to increased interest in the (sub-)GeV mass range \cite{Boehm:2003hm,Fayet:2004bw,Izaguirre:2014bca,Izaguirre:2015yja,Essig:2017kqs,Xu:2018efh}, where many of the strongest constraints can be evaded. One such class of models is a dark sector charged under a new secluded $U(1)_D$ symmetry that is only coupled to  SM particles via kinetic mixing of the associated gauge boson  with the SM hypercharge boson \cite{ArkaniHamed:2008qn,Foot:2014uba,Darme:2017glc,Dutra:2018gmv,Darme:2018jmx}.
\par
While such secluded DM scenarios have been investigated extensively in the past, in this article we study a model where an extra vector-like fermion $\chi$ charged under a gauged $U(1)_{L_\mu-L_\tau}$  symmetry is added to the spectrum. This model  differs crucially from the secluded scenario  in the gauge couplings to second and third generation leptons of the gauge boson associated with the new symmetry. Such a setup is automatically anomaly-free \cite{Foot:1990mn,He:1990pn,He:1991qd} and embeddable into a larger symmetry group $G_{L_\mu-L_\tau}$ \cite{Bauer:2018onh}. In the literature DM charged under a $U(1)_{L_\mu-L_\tau}$  symmetry has been studied for heavy (weak-scale) WIMPs \cite{Baek:2008nz,Baek:2015fea,Arcadi:2018tly,Bauer:2018egk,Altmannshofer:2016jzy,Biswas:2016yan,Biswas:2016yan}.  However, in this article we explore the MeV mass range.
The purpose of this work is to show  that there the $(g-2)_\mu$ anomaly \cite{Baek:2001kca,Ma:2001md,Fayet:2007ua,Altmannshofer:2014pba} and the observed DM relic abundance $\Omega_\text{DM}h^2$ can \emph{simultaneously} be explained. This is not possible in the simple secluded dark sector scenario.
\par
 Recently, a similar scenario has been considered in \cite{Kahn:2018cqs}. However, our work differs in three crucial aspects: $i)$ We fully take into account unavoidable loop-induced kinetic mixing $-\epsilon'/2\, \hat F_{\mu\nu}\hat X^{\mu\nu}$ of the hidden photon with the SM hypercharge boson \cite{Holdom:1985ag}.
Not only does the mixing have important observational consequences  (especially in the hadronic sector), but setting it to zero also requires quite some amount of fine tuning in order to exactly cancel all higher-order loop-contributions.
$ii)$ While the scenario explaining the EDGES anomaly presented in \cite{Kahn:2018cqs} requires  a charge hierarchy of at least $\mathcal{O}(10^2)$, we do not impose any charge hierarchies larger than already present in the SM (i.e. $\mathcal{O}(10)$). Instead we focus on the case of $Q_\chi=1$. 
$iii)$ We include and calculate the full set of constraints on the associated hidden photon $A'$ presented in \cite{Bauer:2018onh}.\par
In the remainder of this article we will first introduce the model, then discuss the various dark matter and hidden photon constraints, before we will present our results and conclude.

%%%%%%%%%%%%%%%%%%%%%%%%%%%%%%%
\section{The Model}\label{sec:model}
%%%%%%%%%%%%%%%%%%%%%%%%%%%%%%%

We extend the minimal $U(1)_{L_\mu-L_\tau}$  model presented in \cite{Bauer:2018onh} by a vector-like fermion $\chi$ with mass $m_\chi$\footnote{In this article we are not studying the scalar breaking of the  $U(1)_{L_\mu-L_\tau}$  and  treat $m_\chi$ and $M_{A'}$ as free parameters.} and $\mu\tau$-charge $Q_\chi$, given by the Lagrangian 
\begin{align}
\mathcal{L}_\chi=-g_{\mu\tau}\, Q_\chi \, \bar \chi \gamma_\mu \chi \hat X^\mu - m_\chi \bar \chi \chi  \,,
\end{align}
where $g_{\mu\tau}$ denotes the $U(1)_{L_\mu-L_\tau}$ gauge coupling and $\hat X$ the associated boson in the gauge basis.
As $\chi$ is vector-like and only carries $U(1)_{L_\mu-L_\tau}$ charge, it does not contribute to the   kinetic mixing  $\epsilon_{\mu\tau}(q^2)$ induced at one-loop so that it is the same as in the minimal case  (cf. Eq. (9) of \cite{Bauer:2018onh}).
We follow Appendix A of \cite{Bauer:2018onh} to canonically normalize the kinetic terms and rotate to the mass basis of the hidden photon denoted by $A'$.
As we are only interested in the light regime $M_{A'} < M_Z$, the hidden photon will only decay into fermionic final states with partial decay widths
\begin{align}\label{eq:Apwidth}
\Gamma_{A'\to f \bar f} =M_{A'}\frac{g_{\mu\tau}^2Q_f^2N_c^f}{12\pi}\sqrt{1-\frac{4m_f^2}{M_{A'}^2}}\Big(1+2 \frac{m_f^2}{M_{A'}^2}\Big)\,.
\end{align}
Here $m_f, Q_f$ and $N_c^f$ denote the fermion mass, $\mu\tau$-charge and number of colors. For the left-handed massless neutrinos one has to set $m_f=0$ and divide by a factor of $2$. For electrons one has to replace  $g_{\mu\tau}Q_f\to \epsilon_{\mu\tau}(q^2) e$. As the $A'$ couples to hadrons only via kinetic mixing we make use of the measured ratio of hadronic to muonic final states $R(s)=\sigma(e^+e^-\to \text{had})/\sigma(e^+e^-\to \mu^+\mu^-)$ \cite{Ezhela:2003pp,Patrignani:2016xqp} to parametrize the hadronic partial width as
  \begin{align}
\Gamma_{A'\to \text{had}}=\epsilon_{\mu\tau}(M_{A'}^2)^2\,\Gamma_{\gamma^\ast\to \mu^+\mu^-}R(M_{A'}^2)\,,
\end{align}
where $\Gamma_{\gamma^\ast\to \mu^+\mu^-}$ is the partial width of a virtual photon $\gamma^*$ of mass $M_A'$. 
Compared to the hidden photon width $\Gamma_{A',0}$  of the minimal model studied in \cite{Bauer:2018onh}, the total width  is increased by the $\chi \bar\chi$ contribution $\Gamma_{A'} = \Gamma_{A',0} + \Gamma_{A'\to\chi\bar\chi}$.
The additional $\chi \bar\chi$ channel will also increase the invisible branching fraction of the $A'$, making this scenario more sensitive to  invisible searches.\par 

As we are exploring the near resonance region $m_\chi \lesssim M_{A'}/2$  we have to perform the full thermal integral when calculating  the thermally averaged annihilation cross section $\langle\sigma v\rangle$ of the DM particle $\chi$ \cite{Griest:1990kh}. This is found to be \cite{Gondolo:1990dk}
\begin{eqnarray}
\langle \sigma v \rangle_\text{CM} &=& \frac{x}{2\,[K_1^2(x)+K_2^2(x)]} \nonumber \\
&&\times \int_{2}^\infty\text{d}z \,\sigma(z^2m_\chi^2)(z^2-4)z^2\,K_1(zx) \,,
\end{eqnarray}
where $x=m_\chi/T$, $z=\sqrt{s}/m_\chi$ and $K_n(x)$ are the modified Bessel functions of the second kind. \\
The cross section of the process $\chi\bar\chi\to f\bar f$ can be expressed as \footnote{Contributions to the cross section from $Z$-mediation are suppressed by at least a factor of $\epsilon_{\mu\tau}/g_{\mu\tau}\ M^2_{A'}/M^2_{Z}$ and may safely be neglected below the $Z$-resonance.}
\begin{eqnarray}
\sigma_{A'}(s) &=& \frac{g_{\chi A'}^2g_{fA'}^2 \,N_c^f}{12\,\pi} \sqrt{\frac{s-4m_f^2}{s-4m_\chi^2}} \nonumber \\
&& \times\ \frac{s^2+2(m_\chi^2+m_f^2)s+4m_\chi^2m_f^2}{s\left((s-M_{A'}^2)^2+\Gamma_{A'}^2M_{A'}^2\right)}\,,
\end{eqnarray}
with $g_{\chi A'}$ and $g_{f A'}$  denoting the couplings of the $A'$ to the particles $\chi$ and $f$ in the mass basis. 
\\

\paragraph{Relic density}

We have implemented our model in \texttt{Feynrules} \cite{Alloul:2013bka} and solve the Boltzmann equation
\begin{equation}
 \dot{n}_\chi + 3 H n_\chi = - \frac{\langle\sigma v  \rangle}{2} (n^2_\chi-n^2_{\chi,\text{eq}})\label{eq:BEnumber}
\end{equation}
tracking the time-evolution of the DM number density $n_\chi$ numerically with \texttt{MadDM} v.3.0 \cite{Ambrogi:2018jqj}.
This way we obtain the freeze-out  temperature $x_f$ at chemical decoupling and relic abundance $\Omega_{\chi}h^2$ of the DM particle $\chi$.
\\

\paragraph{Kinetic decoupling}

After chemical decoupling the DM is still kept in local thermal equilibrium with the SM plasma by elastic scattering processes and tracks the plasma temperature $T$ \cite{Bringmann:2009vf}. Only after elastic scattering has become inefficient the two sectors are kinetically decoupled and the temperature of the DM gas will evolve as that of non-relativistic matter. In \cite{Bringmann:2006mu} the DM temperature evolution $T_\chi(T)$ has been derived  to be
\begin{equation}
\frac{T_\chi}{T} =   1 - \frac{z^{1/(n+2)}e^z}{n+2} \, \Gamma\left[\frac{-1}{n+2}, z\right] \Bigg|_{z=\frac{a}{n+2}\left(\frac{T}{m_\chi}\right)^{n+2}} \,,
\end{equation}
where $a$ is a constant proportional to the leading order expansion  coefficient $c_n$ of the  zero-momentum elastic scattering amplitude $|\mathcal{M}|^2_{t=0} \propto c_n (\omega/m_\chi)^n$ and $n$ the scaling exponent.
Finally, one can use the relation  
\begin{equation}
\frac{T_\text{kd}}{m_\chi} = \left( \left( \frac{a}{n+2}\right)^{1/(n+2)}\ \Gamma\left[\frac{n+1}{n+2}\right]\right)^{-1} \, \label{eq:Tkd}
\end{equation}
to obtain the decoupling temperature $T_\text{kd}$.

%%%%%%%%%%%%%%%%%%%%%%%%%%%%%%%
\section{Dark Matter Constraints}\label{sec:cosmo_constr}
%%%%%%%%%%%%%%%%%%%%%%%%%%%%%%%{}

\begin{figure*}[th!]
%\hspace{-10.cm}
\begin{center}
\includegraphics[width=0.5\textwidth]{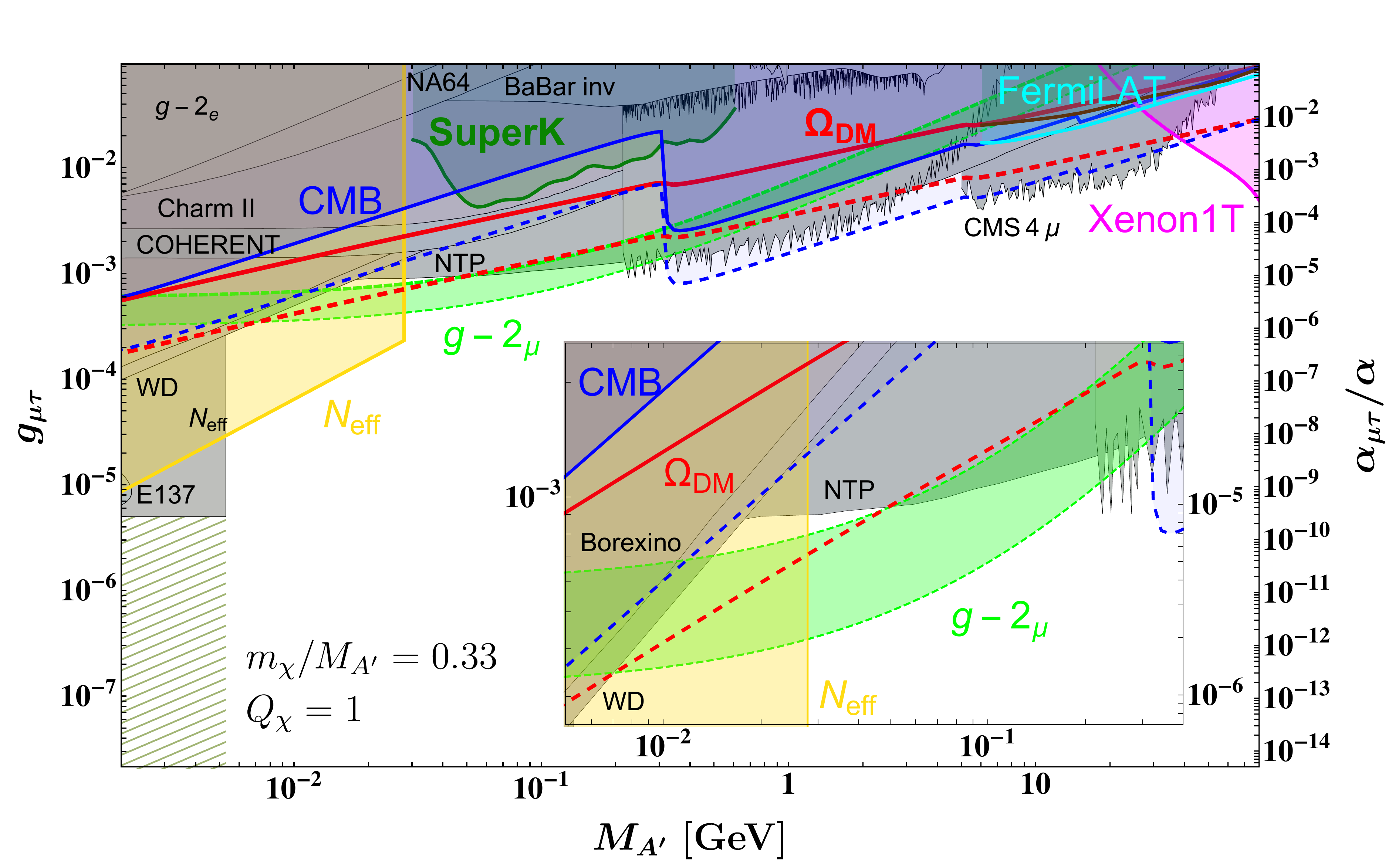}%
\includegraphics[width=0.5\textwidth]{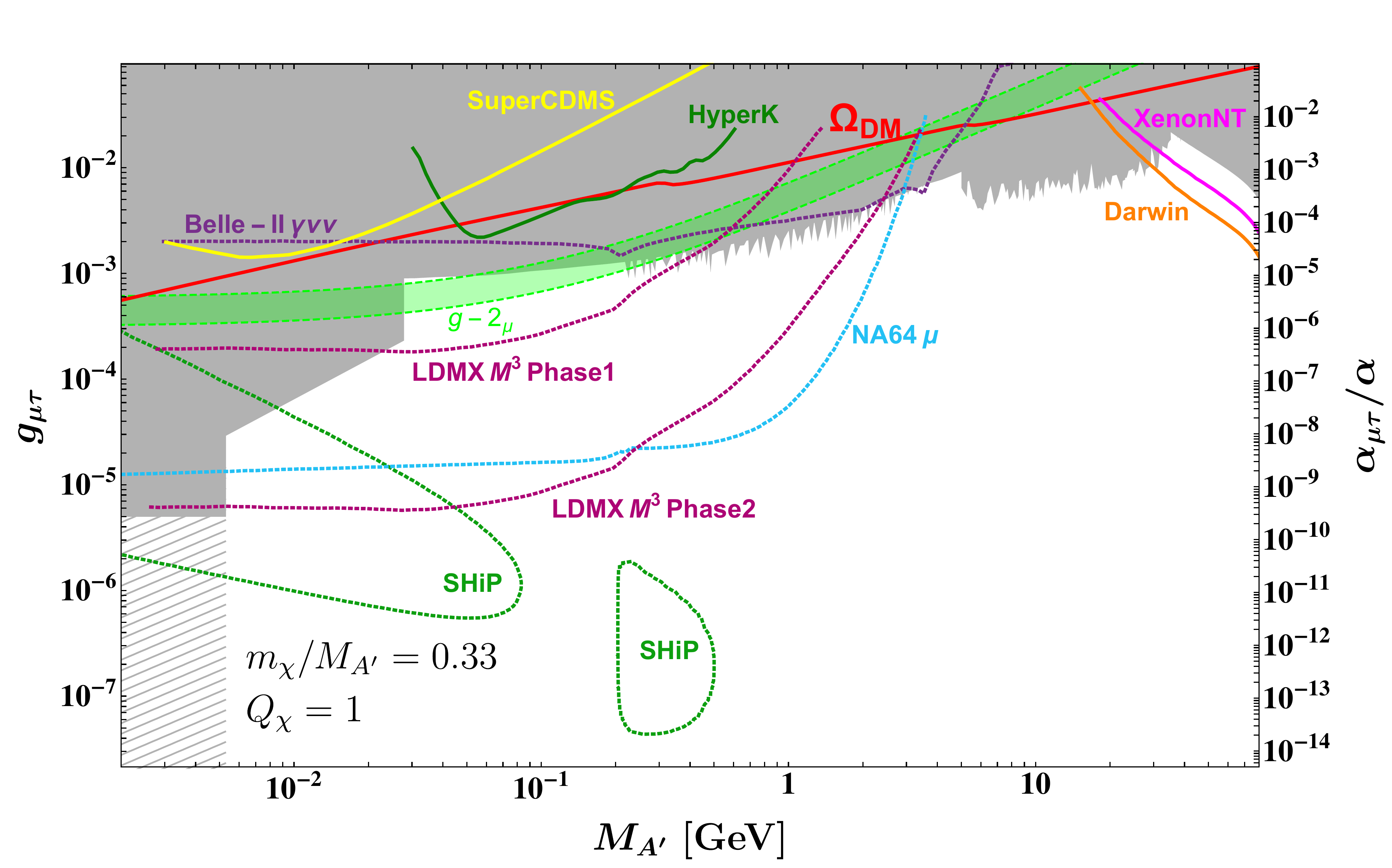}
\includegraphics[width=0.5\textwidth]{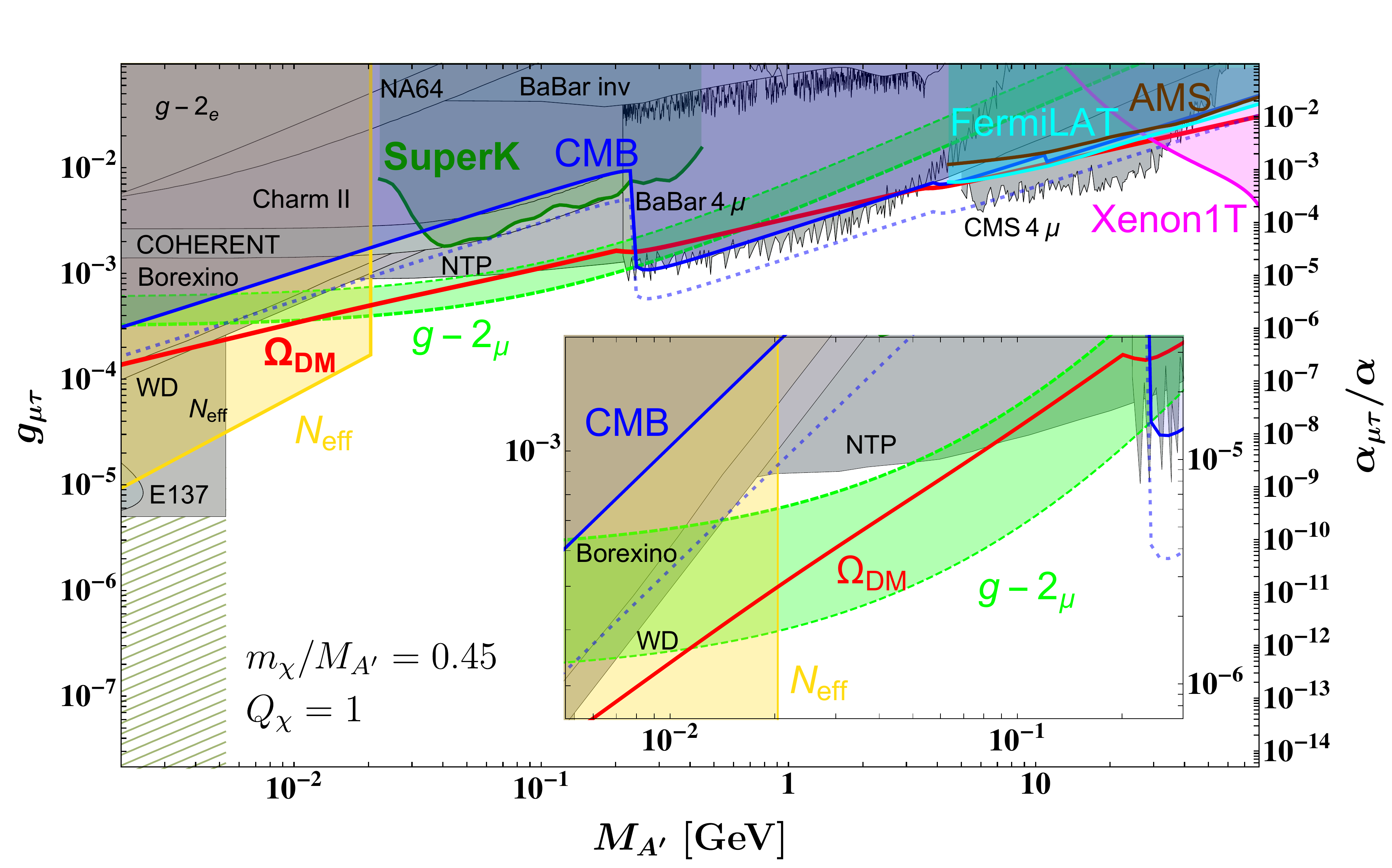}%
\includegraphics[width=0.5\textwidth]{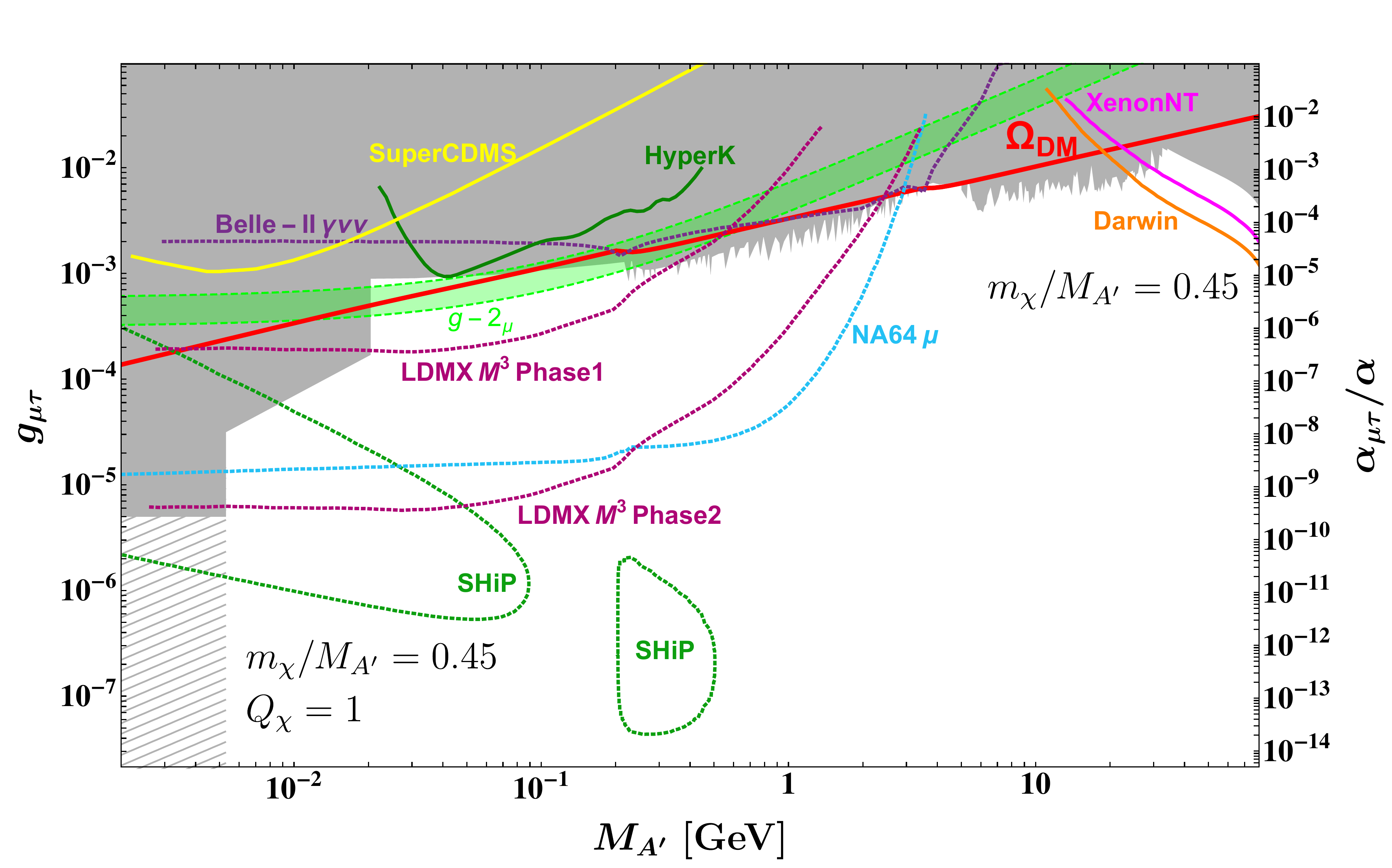}
%\vspace{-.8cm}
\end{center}
\caption{\label{fig:limits}  Current (left panels) and future (right panels) constraints on the combined parameter space of a $U(1)_{L_{\mu}-L_\tau}$ gauge boson and vector-like fermion of charge $Q_\chi=1$ for $m_\chi/M_{A'}=0.33$ (upper panels) and $m_\chi/M_{A'}=0.45$ (lower panels).}\vspace{-.4cm}
\end{figure*}

In this section we discuss the various constraints on the DM parameter space shown in \Cref{fig:limits,fig:limitsQN}. 
\\

\paragraph{CMB}\label{sec:cmb}

 A significant increase in the post-recombination ionization would be
visible as extra free electrons and photons broadening the last scattering
surface of the CMB photons, thus modifying their temperature and polarization power
spectra \cite{Padmanabhan:2005es}. The amount of additional
energy released per unit volume \cite{Slatyer:2012yq} is quantized as
\begin{align}\label{eq:Injection}
\frac{\mathrm{d}E}{\mathrm{d}t\mathrm{d}V}(z) = 2 g \rho_\text{crit}^2
  c^2 \Omega_c (1+z)^6 P_\text{ann}(z) \,,
  \end{align}
with the model-dependent annihilation parameter $P_{\mathrm{ann}}(z) = f(z) {\langle \sigma v \rangle}/{m_\chi}$. 
The efficiency factor $f(z)$ characterizing  the fraction of rest mass energy released into the gas \cite{Slatyer:2009yq,Slatyer:2012yq} depends \textit{a priori} on the redshift $z$.
However, $f(z)$ is to good approximation independent of $z$ and has been calculated for electron and photon final states in \cite{Slatyer:2015jla}. Using these, a mass-dependent effective efficiency factor quantizing the amount of energy proceeding into photons and electrons  can be calculated,
\begin{eqnarray}
f_{\mathrm{eff}}(m_\chi) = \frac{1}{2m_\chi} \int_0^{m_\chi} E \mathrm{d}E \Big[& 2 f^{e^+e^-}_\mathrm{eff}(E) \left(\frac{\mathrm{d}N}{\mathrm{d}E}\right)_{e^+} \nonumber \\
&+ f^{\gamma}_\mathrm{eff}(E)  \left(\frac{\mathrm{d}N}{\mathrm{d}E}\right)_{\gamma}\Big] \,.
\end{eqnarray}
The annihilation parameter  has recently been constrained to  $P_{\mathrm{ann}} < 3.5\times10^{-28} \text{cm}^3 s^{-1}$ GeV by Planck results \cite{Aghanim:2018eyx}, constituting an  improvement of about 17\% over previous results \cite{Ade:2015xua}. 
For $m_\chi>5$ GeV we have used this bound together with the effective efficiency factors $f_\text{eff}$ provided in  \cite{Slatyer:2015jla} to robustly constrain  $\langle\sigma v\rangle$. For $m_\chi<5$ GeV we have applied the more conservative estimate of $\langle\sigma v\rangle/m_\chi < 5.1 \times10^{-27}  \text{cm}^3 s^{-1} \text {GeV}$ (rescaled by the 17 \% improvement from Planck) derived in \cite{Leane:2018kjk}. \\

\paragraph{Big Bang Nucleosynthesis (BBN)}\label{sec:bbn}

As  $\chi$ couples to neutrinos via the $A'$ it will reheat the neutrino gas once it becomes non-relativistic at a temperature $T\sim m_\chi$  \cite{Boehm:2012gr}. This increases the effective number of neutrinos $N_\text{eff}$ and thus the neutrino-to-photon temperature ratio $T_\nu/T_\gamma$ compared to the SM,  resulting in a higher expansion rate $H$ of the universe. If the reheating happens during BBN, this manifests in an elevated helium relic abundance $Y_p$ and deuterium-to-hydrogen ratio D/H in the late universe.\par
These deviations  from their respective SM predictions have been confronted with Planck results for light Dirac fermions in \cite{Boehm:2013jpa,Nollett:2014lwa}. The analysis of \cite{Nollett:2014lwa} excludes extra fermions $\chi$ with masses below $m_\chi^\text{BBN} = 9.28$ MeV
assuming that $\chi$ has been in thermal equilibrium with the neutrinos throughout BBN. \par
This constraint is not applicable anymore once the DM particle starts to decouple kinetically from the neutrino gas at BBN temperatures of $T^\text{BBN} \approx 1$ MeV. Therefore, we have scanned the decoupling temperature $T_\text{kd}$  of $\chi$ as a function of the coupling $g_{\mu\tau}$ by use of \Cref{eq:Tkd} and only show the limit where $T_\text{kd}(g_{\mu\tau}) \leq T^\text{BBN}$.
\\

\paragraph{Dwarf galaxies}

Dwarf spheroidal galaxies in our local group exhibit rather significant DM densities \cite{Mateo:1998wg}. Given their local DM densities $\rho_\text{DM}(\boldsymbol r)$ the flux of gamma ray photons  observed in a solid angle $\Delta\Omega$ at Fermi-LAT is given by \cite{Ackermann:2015zua}
\begin{align}
\phi_s(\Delta \Omega) = &\frac{1}{4\pi } \frac{\langle \sigma v \rangle}{2m_\text{DM}^2}  \int_{E_\text{min}}^{E_\text{max}}  \frac{\mathrm{d}N_\gamma}{\mathrm{d}E_\gamma} \ \mathrm{d}E_\gamma \nonumber  \\
& \qquad \qquad \times \int_{\Delta\Omega}\int_{l.o.s} \rho_\text{DM}(\boldsymbol r)^2 \mathrm{d}l \mathrm{d}\Omega'\,.
\end{align}
The second term is the line-of-sight integral through the DM distribution, the so-called $J$-factor, and only depends on astrophysics. Hence, provided knowledge of the differential photon yield per DM annihilation d$N_\gamma$/d$E_\gamma$, one can place limits on $\langle \sigma v \rangle$ from DM annihilation into the different channels in dwarf galaxies \cite{Ackermann:2015zua,Fermi-LAT:2016uux}.
\\

\paragraph{Cosmic ray positron flux}

The cosmic ray positron fraction has been measured with the AMS-02 detector \cite{Aguilar:2014mma,Accardo:2014lma}. 
 Decays of DM particles source additional  positron injection described by the source term \cite{Bergstrom:2013jra}
\begin{equation}
Q_\chi = \frac{1}{2} \langle\sigma v\rangle \left(\frac{\rho_\chi}{m_\chi}\right)^2 \sum_f \frac{\mathrm{d}N_f}{\mathrm{d}E}\,,
\end{equation}
where ${\mathrm{d}N_f}/{\mathrm{d}E}$ denotes the produced positron spectrum from a $f\bar f$ final state. The AMS-02 data has been used in \cite{Leane:2018kjk} to set an upper limit on $ \langle\sigma v\rangle$  for different final states, with the most stringent one coming from the muon channel for $U(1)_{L_\mu-L_\tau}$.
\\

\paragraph{Neutrino production}

Neutrino detectors with a good energy resolution and low threshold are sensitive to the neutrino flux generated in annihilations of MeV-scale DM. The Super-Kamiokande water Cherenkov detector was used to set limits on the thermal cross section \cite{PalomaresRuiz:2007eu}. In a recent study the sensitivity of the planned Hyper-Kamiokande experiment has been explored \cite{Campo:2018dfh}.\\

\paragraph{Electron scattering}

Kinetic mixing of the hidden photon with the SM photon leads to a non-zero elastic DM-electron scattering cross section \cite{Essig:2015cda}
\begin{align}
\bar\sigma_e = \frac{16\, \pi\, \mu_{\chi e}^2 \alpha \epsilon_{\mu\tau}^2(q^2) \alpha_{\mu\tau}}{(m_{A'}^2+\alpha^2 m_e^2)^2} \,,
\end{align}
 where $\mu_{\chi e}$ denotes the electron-DM reduced mass.

SuperCDMS  \cite{Agnese:2016cpb}  can search for DM-electron scattering events to constrain the DM-electron scattering cross section. 
\\

\paragraph{Nuclear scattering}

Similar to electrons, kinetic mixing  induces a non-zero elastic DM-nucleon scattering \cite{Evans:2017kti}
\begin{align}
\sigma_N = \frac{1}{\pi}g_{\mu\tau}^2\,\epsilon_{\mu\tau}^2(q^2) \mu_{\chi N}^2 \left| \frac{f^{(A')}_N}{M_{A'}^2}-\frac{s_W\,f^{(Z)}_N }{M_Z^2-M_{A'}^2}\right|^2 \,,
\end{align}
with the reduced DM-nucleon mass $\mu_{\chi N}$ and 
\begin{align}
f^{(X)}_N = \frac{1}{A}\left(Z(2 g_{uX}+ g_{dX})+(A-Z)( g_{uX}+ 2g_{dX})\right) \,.
\end{align}
Here, $A$ and $Z$ refer to mass and atomic number of the nucleus and $g_{qX}=g_{qX_L}+g_{qX_R}$ denotes the sum of the chiral couplings  of the quark $q$ to the boson $X$ \cite{Arcadi:2018tly}. 
The current best limit on the DM-nucleon scattering cross section comes from the 1 t $ \times$ yr exposure dataset of the XENON1T experiment \cite{Aprile:2018dbl}.  
\par
The proposed DARWIN experiment will be able to probe even smaller cross sections  \cite{Aalbers:2016jon}.

%%% ++++++++ 
%%% ++++++++ 

%%%%%%%%%%%%%%%%%%%%%%%%%%%%%%%
\section{Hidden Photon constraints}\label{sec:hp_constr}
%%%%%%%%%%%%%%%%%%%%%%%%%%%%%%%

We briefly summarize the most important hidden photon constraints, which are discussed  in detail in our previous article  \cite{Bauer:2018onh}.
\\

As the $A'$  couples to neutrinos and electrons it contributes to the in-medium plasmon decay $\gamma^*\to\nu\nu$. This can be constrained from the good agreement of SM white dwarf cooling with observations \cite{Dreiner:2013tja}.
Furthermore, the agreement of the measured cross section of  neutrino trident production  $\nu N\to \nu N \mu^+\mu^-$ with its SM prediction  \cite{Geiregat:1990gz,Mishra:1991bv,Adams:1998yf} puts strong limits on the $A'$ parameter space \cite{Altmannshofer:2014pba}. Likewise,  the measured  elastic $\nu-e$ scattering cross section  with the Borexino detector constrains  $g_{\mu\tau}$ \cite{Kaneta:2016uyt}. \par
The BaBar  \cite{TheBABAR:2016rlg} and CMS \cite{Sirunyan:2018nnz} searches for four-muon final states put quite stringent constraints on the hidden photon of a muonic force.
The proposed muon runs of both NA64 \cite{Gninenko:2014pea,Gninenko:2018tlp} and LDMX \cite{Kahn:2018cqs,Berlin:2018bsc} could  exclude large parts of the $A'$ parameter space from a missing energy search. Similarly, SHiP \cite{Anelli:2015pba,Alekhin:2015byh} will be able to test very small couplings from a search for the visible $A'$ decays.\par
Finally, it has been shown \cite{Ma:2001md} that a $U(1)_{L_\mu-L_\tau}$ boson can potentially explain the observed shift $\Delta a_\mu = (2.87 \pm 0.80) \times 10^{-9}$   \cite{Bennett:2006fi,Davier:2010nc,Hagiwara:2011af} of the muon anomalous moment.

%%%%%%%%%%%%%%%%%%%%%%%%%%%%%%%
\section{Results}\label{sec:results}
%%%%%%%%%%%%%%%%%%%%%%%%%%%%%%%

\begin{figure}[t!]
%\hspace{-10.cm}
\begin{center}
\includegraphics[width=0.5\textwidth]{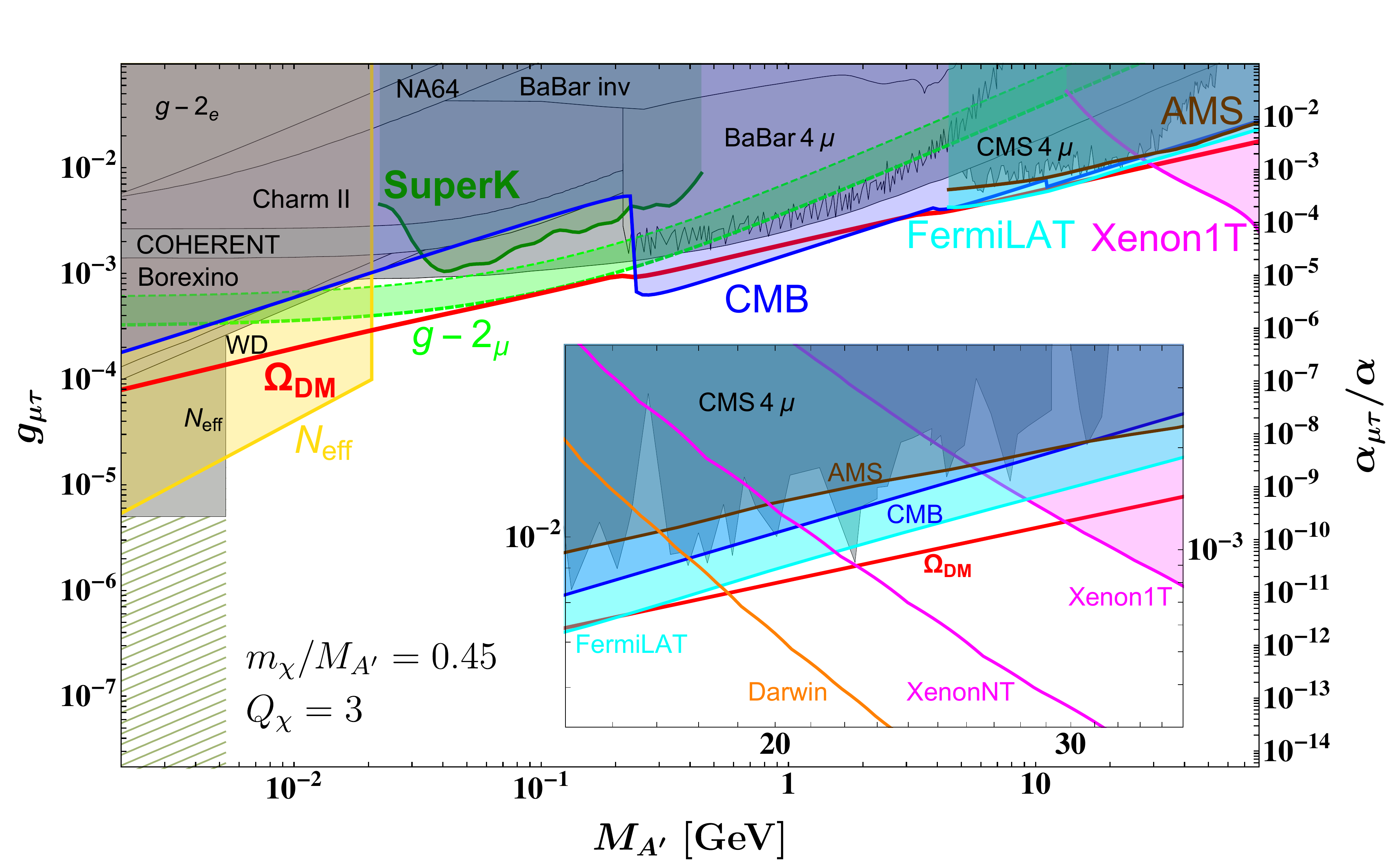}
\end{center}
\caption{\label{fig:limitsQN}  Same as lower left panel of \Cref{fig:limits}, but with $Q_\chi=3$.}\vspace{-.4cm}
\end{figure}

In the following we discuss our results summarized in  \Cref{fig:limits,fig:limitsQN}.\\

 In \Cref{fig:limits} we compare the standard benchmark  (SB) scenario of  ${m_\chi}/{M_{A'}}=0.33$ (upper panels) to a near-resonance (NR) scenario  with ${m_\chi}/{M_{A'}}=0.45$ (lower panels)  for a DM charge of $Q_\chi =1$.
The grayscale contours (dotted lines) in the left (right) panels show current (projected) hidden photon constraints. The constraints from white dwarf cooling (WD), neutrino trident production (NTP), and elastic neutrino scattering experiments (Borexino, COHERENT, CharmII) are completely insensitive to the DM $\chi$. Constraints from visible decay searches  (e.g. BaBar/CMS 4$\mu$, SHiP)  are generally weakened due to the reduced visible branching fraction of the $A'$ compared to the minimal case of \cite{Bauer:2018onh}. Conversely, constraints from invisible searches (e.g. NA64 $\mu$, LDMX M$^3$) become more stringent. \par
The colored contours (solid lines) in the left (right) panels represent current (projected) DM limits. While at low masses the most stringent limit is the one from $N_\text{eff}$ at BBN (yellow), the high-mass region is most  constrained by the Xenon1T (magenta) limit on kinetic-mixing-induced DM-nucleon scattering. In the intermediate-mass region a combination of measurements of gamma rays at FermiLAT (cyan), positron fluxes at AMS-02 (brown) and energy injection into the CMB (blue) at Planck yield the most constraining limit on $\chi$ and the $4\mu$ searches at BaBar and CMS on $A'$.\par
The red curve represents points where $\Omega_\chi = \Omega_\text{DM}$.
Points above the red curve correspond to $\chi$ being a subcomponent of DM. 
First, we note that in the SB an explanation of  $\Omega_\text{DM}$ is entirely ruled out for unit charge. Only by increasing the charge to $Q_\chi=10$ (for which the relic density and the CMB bound are shown by the dashed red and blue curves in the upper left panel of \Cref{fig:limits}) the DM relic abundance can be explained for $28 \lesssim M_{A'}\lesssim 50$ MeV.\par
In the NR case, however, we can accommodate a simultaneous explanation of  $\Omega_\text{DM}$ and the $(g-2)_\mu$ anomaly for $20 \lesssim M_{A'} \lesssim 85$ MeV  with unit charge\footnote{We note that for mass ratios ${m_\chi}/{M_{A'}}<{m_\chi^\text{BBN}}/{2 m_{\mu}} \approx 0.044$ an explanation of $(g-2)_\mu$ is excluded by the $N_\text{eff}$ bound.} (see inset plot in the lower left panel of \Cref{fig:limits}). This region is not excluded by the CMB bound, in particular it is also unaffected by the theoretical maximum CMB bound (dotted blue in the lower left panel of \Cref{fig:limits}) obtained from assuming that visible annihilation products inject all of their energy into the CMB. This low-mass region will be testable at  the future muon beam experiments NA64$\mu$ \cite{Gninenko:2018tlp} and LDMX M$^3$ \cite{Kahn:2018cqs}. \\

While excluded for $Q_\chi=1$ by the recent CMS $4\mu$ search \cite{Sirunyan:2018nnz}, there is still a window for a high-mass explanation of $\Omega_\text{DM}$ at  $16.5 \lesssim M_{A'} \lesssim 30$ GeV\footnote{In the intermediate-mass region of $3 \lesssim M_{A'} \lesssim 5$ GeV $\Omega_\text{DM}$ is also not yet excluded. However, this is most probably an artifact of our conservative estimate of the CMB bound below  $m_\chi=5$ GeV \cite{Leane:2018kjk}.}  for $Q_\chi=3$ or higher (see \Cref{fig:limitsQN}). Interestingly, this region grows upon increasing the ratio  $m_\chi/M_{A'}$ up to $m_\chi = M_{A'}/2$. The reason for this region reaching below the model-independent lower mass bound of $m_\chi \gtrsim 20$ GeV derived in \cite{Leane:2018kjk} is twofold: first, $\chi$ has a sizable invisible annihilation branching fraction into neutrinos, which do not leave any imprint on the CMB in this mass range \cite{Leane:2018kjk}.  Second, our FermiLAT bound is conservative in the sense that we have used the limit $\langle\sigma v\rangle$ obtained for 100\% branching fraction into taus. From the inset plot of \Cref{fig:limitsQN} we can see that this high-mass region can be tested by the future XenonNT \cite{Aprile:2015uzo} and Darwin experiments \cite{Aalbers:2016jon}. \\

Summing up, we have seen that while the SB scenario is not capable of explaining DM with unit charges, in the NR case $\Omega_\text{DM}h^2$ and $(g-2)_
\mu$ can simultaneously be explained and tested at future muon beam dump facilities.
\\

%%% ++++++++ 
%%% ++++++++ 

%\end{document}

\begin{acknowledgments}
The author is grateful to  Giorgio Arcadi, Martin Bauer, Bj\"{o}rn-Malte Sch\"{a}fer, Sebastian Schenk and especially J\"{o}rg J\"{a}ckel for many helpful and inspiring discussions. The author thanks Michael Russell for valuable comments on this manuscript. The author also acknowledges support from the DFG via the GRK 1940 "Particle Physics beyond the Standard Model". 
\end{acknowledgments}

\end{document}